\documentclass[aps,pra,twocolumn,showpacs,superscriptaddress]{revtex4}

\usepackage{graphicx} 

\begin{document}

\title
{Photonic Band Gap in the Triangular Lattice of BEC vortices}

\author{M.E. Ta\c{s}g{\i}n}
\affiliation{Department of Physics, Bilkent University, 06800 Bilkent,
  Ankara, Turkey}
\author{\"{O}. E. M\"{u}stecapl{\i}o\u{g}lu}
\affiliation{Department of Physics, Ko\c{c} University, 34450
Sar{\i}yer, Istanbul, Turkey}
\author{M.\"{O}. Oktel}
\affiliation{Department of Physics, Bilkent University, 06800
Bilkent,
  Ankara, Turkey}

\date{\today}

\begin{abstract}
We investigate the photonic bands of an atomic Bose-Einstein
condensate with a triangular vortex lattice. Index contrast
between the vortex cores and the bulk of the condensate is
achieved through the enhancement of the index via atomic
coherence. Frequency dependent dielectric function is used in the
calculations of the bands, resulting in photonic band gap widths
of a few MHz.
\end{abstract}


\maketitle

\section{Introduction} \label{sec:intro}

A rotating Bose-Einstein Condensate (BEC) manifests the formation
of vortices after a critical rotation frequency. Furthermore,
constituent vortices exhibit a periodic structure, which is
generally a triangular lattice \cite{1,2,3,baym}. Even near the
borders of the condensate, lattice distortion is small
\cite{baym}. From the theoretical point of view, a rapidly
rotating BEC can be treated analytically and the density is found
to be the product of a slowly varying function and a periodic
function \cite{3,baym}.

Usual imaging of the vortices is carried on during the ballistical
expansion of the condensate \cite{1,2,5,6,7}. An in-situ imaging
was accomplished recently, by visualizing the 2D image of the
lattice along the rotation axis, while condensate is in the trap
\cite{8}.

In the BEC experiments, rotation frequency is not directly
measured, but deduced from the change in the aspect ratio of the
cloud \cite{1,2,5,6,7}. However, we recently proposed a method,
which is based on the reflection through the directional pseudo
photonic band gaps \cite{mustecap}. Photonic band gaps in a
triangular vortex lattice are obtained through the enhancement of
the refractive index via quantum coherence \cite{scully,quaopt} so
that sufficient index difference is generated between the vortex
cores and the bulk of the condensate. Without an index enhancement
scheme, a usual rotating BEC with a vortex lattice cannot exhibit
high enough index contrast to obtain photonic band gaps. BECs are
rather dilute, and being an atomic gaseous medium, they exhibit
dispersion only in highly absorptive regimes. When the ground and
excited state are coupled to other auxiliary levels, however,
absorbtion in the resonant transition of the probe beam can be
cancelled. This is due to the quantum interference of various
absorbtion paths. This way, one can benefit from large dispersion
at an atomic resonance, without absorption.

Utilizing the index enhancement scheme, triangular lattice of BEC
vortices can generate both directional and complete photonic band
gaps. Directional pseudo-gaps are also called as partial or stop
gaps \cite{pulos}. Radiation cannot propagate in certain
directions determined by these pseudo-gaps, but can in others. The
rotation frequency of the condensate can be measured from the
chopping in the reflected or transmitted probe beam at a
directional photonic band gap.

In a previous study, \cite{mustecap} we have demonstrated the
presence of a photonic band gap within the frequency window of
index enhancement. In this paper, we discuss the photonic band
structure for the full frequency regime, extending our work beyond
the index enhancement window.

Though our general examination simply verifies existence of the
band gap within the index enhancement window, beyond this region
electric susceptibility becomes a complex valued function of
frequency, for which definition of photonic band gaps are not
straightforward. In the particular index enhancement scheme we
consider here, there are absorption and gain regimes, where we
have found no photonic band gaps. This allows for selective
stoppage of the probe pulse among the other electromagnetic fields
that are in use for trapping the condensate atoms and for index
enhancement schemes.

Despite the dense literature on photonic crystals, studies of
photonic energy bands which take complex, frequency dependent
dielectric constants into account are rather sparse. Few recent
studies investigate materials with small absorption \cite{krokhin}
and low filling fractions of the dispersive and absorptive
component \cite{sigalas,kuzmiak,li}. More recent works go beyond
these limitations \cite{moroz,tip,lem}. These efforts focus on
understanding the properties of photonic crystals fabricated from
metallic materials, which can have large complete photonic band
gaps in the visible region of the electromagnetic spectrum.
Drude-like models of metallic components within a dielectric host
is used in the modelling of such crystals. Absorption is of
negligible importance within transparency window of the Drude
model which is about half of the plasma frequency. Small, yet
realistic, amounts of absorption hardly changes the band structure
\cite{kuzmiak}. In the regions of appreciable absorption however,
there may be no band gaps \cite{tip}. Our results, for a rotating
atomic BEC, are in agreement with these results for photonic
crystals of metallic materials. We note that, quite recently,
broadband absorptive properties of metallic photonic crystals are
found to be advantageous for various applications \cite{veronis}.
Similarly, gain regime is important for understanding lasing
properties of photonic crystals \cite{jiang}. Index enhancement
schemes we suggest to use for rotating BEC, offer absorptive as
well as gain regimes beyond the index enhancement window,
\cite{scully} associated with lasing without inversion,
\cite{quaopt}.

The understanding of photonic crystals with complex dielectric
constants is not trivial, due to the lack of well-defined group
velocity for complex energy bands \cite{rostovtsev}. In this
paper, we limit ourselves to the determination of the band
structure, and do not discuss the details of beam propagation
beyond the existence of the band gaps.

The paper is arranged as follows. In section \ref{sec:enhencement}
we overview the upper-level microwave scheme, which leads into the
index enhancement with vanishing absorbtion. We introduce the
system parameters and  the resulting dielectric susceptibility. In
section \ref{sec:bands}, we obtain the matrix equations from the
Master Equation of the photonic crystals and illustrate the method
of solutions for the case of frequency dependent, complex
dielectric function. In section. \ref{sec:results}, we present the
resulting photonic bands for two different lattice parameters and
then discuss the properties of the photonic bands. Section
\ref{sec:summary} is a summary of our results.

\section{Dielectric function of the vortex lattice} \label{sec:enhencement}

In this section, we describe the atomic coherence and path
interference effects leading to high index of refraction with vanishing
absorbtion. We review the derivations of Ref. \cite{scully} in a
compact form and describe the physics of the system. We calculate
the real and imaginary parts of the dielectric susceptibility.

\begin{figure}
\includegraphics[scale=0.2]{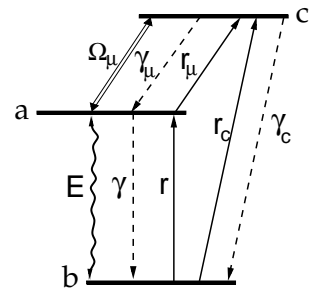}
\caption{Upper-level microwave  scheme for index enhancement
\cite{scully}. Upper two levels $a$ and $c$ are coupled via a
strong microwave field of Rabi frequency $\Omega_{\mu}$. Weak
probe field $E$, of optical frequency $\omega$ is coupled to
levels $a$ and $b$. Decay ($\gamma$) and pump ($r$) rates are
indicated.} \label{fig:levels}
\end{figure}

There are various index enhancement schemes \cite{scully}. Among
them, we specifically consider upper-level microwave scheme as it
leads strong index contrast, though Raman scheme may also be
useful due to its wider frequency window for index enhancement
\cite{scully}. The corresponding level diagram for upper-level
microwave scheme is shown Fig. \ref{fig:levels}. A weak optical
probe field, $E$, of frequency $\omega$ is coupled to the two
levels $a$ and $b$ through electric dipole interaction.  A third
level $c$ is also coupled to level $a$ via a strong resonant
microwave field of Rabi frequency $\Omega_{\mu}$. Coupling level
$a$ to level $c$ allows for different possible paths of
absorption. Destructive quantum interference of these paths may
cancel the absorption of the probe field at a certain frequency
\cite{quaopt}. At the same time, a high refractive index can be
generated by maintaining some population in level $a$ to ensure
high dipole moment for levels $a$ and $b$. Indirect pump
mechanisms are introduced to realize that. Parameters $r_{\mu}$,
$r$ and $r_c$ are the pump rates from level $a$ to $c$, $b$ to $a$
and $b$ to $c$ respectively.

In Fig. \ref{fig:levels}, $\gamma_{\mu}$, $\gamma$ and
$\gamma_{c}$ denote the decay rates (inverse lifetime) of levels
$c$ to $a$, $a$ to $b$ and $c$ to $b$ respectively, due to the
collisions and radiation. We can consider $\gamma_{\mu},\gamma_c
\ll \gamma$, as $\gamma_c$ and $ \gamma_{\mu}$ are for dipole
forbidden and microwave transitions, respectively.

Following Ref. \cite{scully}, we set $r_{\mu}=r=0$, and choose
$r_c=\Omega_{\mu}=\gamma$. We neglect $\gamma_{\mu}$ and
$\gamma_c$. In that case, frequency dependent electric
susceptibility $\chi(\omega)=\chi'(\omega)+i\chi''(\omega)$ is a
complex function of frequency, with its real $\chi'$ and imaginary
parts $\chi''$ are given by \cite{scully}
\begin{eqnarray}
\chi'(\varpi)=\frac{12N\lambda^3}{13\pi^2} \frac{\varpi}
{9-3\varpi^2+4\varpi^4} \label{eq:chi_r}
\\
\chi''(\varpi)=-\frac{3N\lambda^3}{13\pi^2}\frac{-3+2\varpi^2}
{9-3\varpi^2+4\varpi^4} \:, \label{eq:chi_i}
\end{eqnarray}
where $N$ is the number density of atoms and $\lambda$ is the
wavelength of the optical transition $a \rightarrow b$. We define
a dimensionless frequency
\begin{equation}
\varpi=(\omega-\omega_{ab})/\gamma \quad \text{,}
\end{equation}
centered at the resonance frequency $\omega_{ab}$ and scaled with
the decay rate of atomic coherence.

\begin{figure}
\includegraphics[scale=0.4]{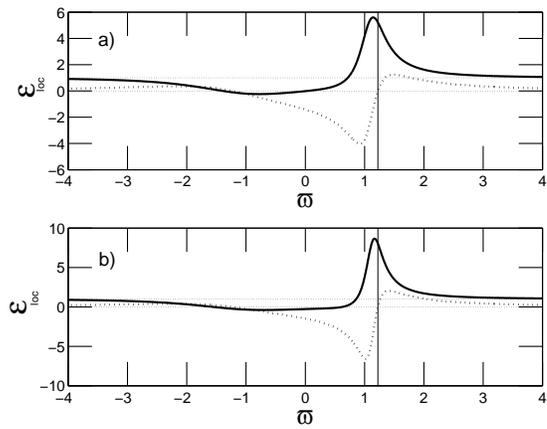}
\caption{Real (solid-line) and imaginary (doted-line) parts of
local dielectric function $\epsilon_{\scriptscriptstyle{{\rm
loc}}}(\omega)$ as a function of scaled frequency
$\varpi=(\omega-\omega_{ab})/\gamma$, for the particle densities
(a) $N=5.5\times10^{20}\text{m}^{-3}$ and (b)
$N=6.6\times10^{20}\text{m}^{-3}$. Vertical solid line indicates
the scaled enhancement frequency $\varpi_0\simeq1.22$, where
$\epsilon_{\scriptscriptstyle{{\rm loc}}}''(\varpi)$ vanishes. (a)
$\epsilon=\epsilon_{\scriptscriptstyle{{\rm loc}}}(\varpi_0)=5.2$
and (b) $\epsilon=\epsilon_{\scriptscriptstyle{{\rm
loc}}}(\varpi_0)=8.0$.} \label{fig:epsilon}
\end{figure}

Susceptibilities, given by Eqs. (\ref{eq:chi_r}) and
(\ref{eq:chi_i}), are for a dilute condensate. In the case of a
dense condensate, first correction to the susceptibility is
equivalent to a local field correction \cite{localfield} in the
form
\begin{equation}
\chi_{\scriptscriptstyle{{ \rm
loc}}}(\varpi)=\frac{\chi(\varpi)}{1-\chi(\varpi)/3} \; ,
\label{eq:localchi}
\end{equation}
Real and imaginary parts of the corresponding dielectric function,
\begin{equation}
\epsilon_{\scriptscriptstyle{{\rm
loc}}}(\varpi)=1+\chi_{\scriptscriptstyle{{\rm loc}}}(\varpi)
\quad , \label{eq:dielectricfreq}
\end{equation}
are plotted in Fig. \ref{fig:epsilon} for a Rubidium 87 gas. The
vertical line indicates the enhancement of the polarization at the
frequency of vanishing absorption. Here, we define
$\Omega_0=2.37\times10^{15}$ Hz as the frequency at which
absorption is zero. We also define
$\varpi_0=(\Omega_0-\omega_{ab})/\gamma\simeq1.22$ as the
corresponding value of the scaled and shifted frequency $\varpi$.
We employ these definitions throughout the paper.

Isotopes of alkali metals are typically used in the BEC
experiments, and we specifically consider the energy levels of
Rubidium. Fine-structure energy levels of Rubidium, which
corresponds to $b$,$a$, and $c$ levels of Fig. \ref{fig:levels},
are $5s_{1/2}$, $5p_{1/2}$, and $6s_{1/2}$, respectively. The
wavelengths of $a-b$ and $a-c$ transitions become $\lambda=794$ nm
and $\lambda_{\mu}=1.32$ $\mu$m. The lifetime of the probe
resonance level ($5p_{1/2}$) is $27$ ns which corresponds to the
decay rate $\gamma=2\pi\times6$ MHz.

In the vicinity of the center of the condensate cloud, the
dielectric function plotted for the peak density in Fig.
\ref{fig:epsilon} can be assumed. When the vortices are present in
this central region however, their spatial profile will influence
the dielectric function. Assuming a dilute thermal gas background
at ultracold temperatures, index enhancement will be influential
on the dense condensate only. Density of the condensate drops
rapidly to zero at the vortex positions. Dielectric constant
within the vortex core can be supposed to be same with that of
vacuum, $\epsilon_0=1$. Width of vortices is in the order of the
coherence length of a condensate, which is given by
$\xi=1/\sqrt{8\pi Na_{sc}}$, where $a_{sc}$ is the s-wave
scattering length of the inter-atomic collisions. The typical
values of the coherence length is a few hundred nanometers and
smaller than the optical wavelength. Spatial modulations on the
dielectric function can be introduced by
$\epsilon=\epsilon_0+(\epsilon_{\scriptscriptstyle{{\rm
loc}}}-\epsilon_0)\rho(\vec{r})$ so that
\begin{equation}
\epsilon_{\scriptscriptstyle{{\rm
loc}}}(\vec{r},\varpi)=1+\rho(\vec{r})\chi_{\scriptscriptstyle{{\rm
loc}}}(\varpi) \: .\label{eq:dielectricfnx}
\end{equation}
Here $\rho(\vec{r})$ stands for the normalized spatial profile of
a vortex, and $\vec{r}$ is the radial distance from the center of
the vortex core. This model dielectric function
(\ref{eq:dielectricfnx}) drops to vacuum value at the cores of the
vortices and recovers its bulk value in a few coherence lengths.

In the band calculations for a triangular lattice of vortices, we
considered a hexagonal Wigner-Seitz unit cell, which contains a
single vortex core at the center of the unit cell. We used
Pad\'{e}'s analytical form, derived in Ref. \cite{pade}, for the
vortex density profile
\begin{equation}
\rho(r)=\frac{r^2(0.3437 + 0.0286r^2)}{1 + 0.3333r^2 + 0.0286r^4}
\: , \label{eq:pade}
\end{equation}
where $r$ is scaled with the coherence length $\xi$. This density
behavior is valid in one unit cell. $\rho(r)$ becomes zero at the
center and goes to $1$ towards the edges of the unit hexagonal
cell. We choose the lattice constant $a$ in terms of the coherence
length to fix the filling factor of the vortices. We used two
different values, $a=10\xi$ and $a=4.5\xi$, in the computations.

In the following sections, we discuss the propagation of probe
beam through the vortex lattice which has a dielectric function
given by Eq. (\ref{eq:dielectricfnx}) and look for possible band
gap formations about the index enhancement frequency.
%
%
%

\section{Calculation of the Photonic Bands} \label{sec:bands}

The stable lattice type for a single component rotating BEC is
triangular \cite{1,2}. Density profile is composed of vortices
distributed periodically and an envelope density profile which
decreases as moving away from the center of the trap. The envelope
is a slowly varying function compared to the periodicity of the
vortices. Radius of the cloud is much greater than the
periodicity, such that there may be few hundred vortices which are
experimentally observable. Moreover, the distortion of the lattice
near the edges of the condensate is small.

In our past work, \cite{mustecap}, we have numerically
investigated the effects of the finite size and imperfections in
the periodicity. We observed that the positions of the gaps are
not strongly effected, despite the occurrence of extra scattering
due to the smooth density envelope over the lattice.

Thus, in this paper we consider an infinite homogenous vortex
lattice and concentrate on the effects of the frequency dependence
of the dielectric function.

A two-dimensional photonic crystal supports only two polarization
modes for the in plane propagation of light \cite{pulos}. If the
magnetic field, $\vec{H}$, is perpendicular to the plane of
periodicity, this mode is called transverse electric (TE). In a TE
mode electric field, $\vec{E}$, is perpendicular to the axis of
vortices. Similarly, the mode with $\vec{E}$ parallel to the
vortex axis is called transverse magnetic mode (TM).

Let's first focus on the TE modes. We take the vortices to be
aligned in the $\hat{z}$ direction, forming a periodic array in
the x-y plane. A generalized eigenvalue equation for
$\vec{H}\parallel\hat{z}$
\begin{equation}
\vec{\nabla} \times \left( \frac{1}{\epsilon(\vec{r},\omega)}
\vec{\nabla} \times \vec{H}(\vec{r}) \right) =
\left(\frac{\omega}{c}\right)^2 \vec{H}(\vec{r})
\label{eq:masterTE}
\end{equation}
is derived by decoupling the Maxwell equations for $\vec{H}$,
after the substitution
$\vec{D}(\vec{r},\omega)=\epsilon(\vec{r},\omega)\vec{E}(\vec{r},\omega)$
\cite{kuzmiak}. Unlike the frequency independent case,
differential operators on the left hand side also depend on
$\omega$. Moreover, differential operator is not Hermitian because
of the imaginary part of the dielectric function
(\ref{eq:dielectricfnx}). This causes eigenfrequencies to be
complex. Since, in general, Eq. (\ref{eq:masterTE}) is not
analytically solvable, we determine the eigenfrequencies
computationally by plane wave expansion.

Using Bloch–-Floquet Theorem \cite{bloch} the magnetic field can be
expressed in terms of the reciprocal lattice vectors $\vec{G}$ as
\begin{equation}
\vec{H}=\sum_{\vec{G}}  H_{\vec{G}} e^{{\rm
i}(\vec{k}+\vec{G})\cdot\vec{r}} \; \hat{z} \; .
\label{eq:expansionH}
\end{equation}
Similarly, inverse dielectric function is expanded as
\begin{equation}
\frac{1}{\epsilon(\vec{r},\omega)}=\sum_{\vec{G}'}
\varepsilon_{\vec{G}'} e^{{\rm i}\vec{G}'\cdot\vec{r}} \quad ,
\label{eq:expansioneps}
\end{equation}
where the Fourier components $\varepsilon_{\vec{G}'}$ are
\begin{equation}
\varepsilon_{\vec{G}'}(\omega)=\frac{1}{A}\int \frac{e^{-{\rm
i}\vec{G}'\cdot\vec{r}}}{\epsilon(\vec{r},\omega)}{\rm d}^2\vec{r}
\quad .
\end{equation}
The integration is carried out over the Wigner-Seitz unit cell of
area $A$.

We substitute the expansions (\ref{eq:expansionH}) and
(\ref{eq:expansioneps}) into the Master equation
(\ref{eq:masterTE}), and obtain the expression
\begin{equation}
\sum_{\vec{G}'} \varepsilon_{(\vec{G}-\vec{G}')}(\omega)
H_{\vec{G}'}  [(\vec{k}+\vec{G})\cdot(\vec{k}+\vec{G}')] =
\left(\frac{\omega}{c} \right)^2 H_{\vec{G}} \: \quad .
\label{eq:hermit}
\end{equation}
We note that, if the dielectric function were real,
$\varepsilon_{\vec{G}-\vec{G}'}^{*}=\varepsilon_{\vec{G}'-\vec{G}}$,
the matrix, represented by equation (\ref{eq:hermit}), would be
real. Eigenfrequencies would also be real. Moreover, due to
inversion symmetry of the unit cell $\epsilon_{\vec{G}}$'s would
be real. However, the presence of complex dielectric function
destroys the Hermiticity. Eigenfrequencies are, in general,
complex.

Real space basis vectors for a triangular lattice are
$\vec{a}_1=a\hat{x}$ and
$\vec{a}_2=a\left(\frac{1}{2}\hat{x}-\frac{\sqrt{3}}{2} \hat{y} \right)$. Corresponding
reciprocal lattice basis vectors are
$\vec{b}_1=k_0 \left(\frac{\sqrt{3}}{2}\hat{x}-\frac{1}{2} \hat{y}\right)$ and
$\vec{b}_2=k_0 \hat{y}$, where magnitude of both vectors are
$k_0=(2/\sqrt{3})(2\pi/a)$. Any lattice point, in the summation,
can be written as $\vec{G}=n_1\vec{b}_1+n_2\vec{b}_2$, where $n_1$
and $n_2$ are all integers. We also denote the fourier components
of inverse dielectric function as
$\varepsilon_{\vec{G}}\equiv\varepsilon_{n_1,n_2}$.

For computational purposes, we limit the number of $\vec{G}$
vectors over which the summation will be carried out. We consider
a parallelogram in the reciprocal space over which $n_1$ and $n_2$
runs through $-N$ to $N$. $N$ is a positive integer. This gives a
$(2N+1)^2\times(2N+1)^2$ matrix of elements
\begin{eqnarray}
\! \! \! \! M_{ij}(\omega)&=
&\varepsilon_{\scriptscriptstyle{\eta_1,\eta_2}}(\omega)
\left[\left(\vec{k}+n_1\vec{b}_1+n_2\vec{b}_2\right)\cdot\left(\vec{k}+n_1'\vec{b}_1+n_2'\vec{b}_2\right)\right]
\nonumber
\\
&-&\left(\frac{\omega}{c}\right)^2 \delta_{ij}
\label{eq:matrix}
\end{eqnarray}
where the dependence of indices are given by
\begin{equation}
i=(2N+1)n_1+n_2 \quad \text{and} \quad j=(2N+1)n_1'+n_2' \; .
\end{equation}
We use the notations
\begin{equation}
\eta_1=n_1-n_1' \quad \text{and} \quad \eta_2=n_2-n_2' \quad .
\label{eq:shorthands}
\end{equation}

Solution of Master equation (\ref{eq:masterTE}) reduces to the
determination of eigenfrequencies $\omega$ for each wave vector
$\vec{k}$. All distinct values of $\omega$ are obtained by
choosing $\vec{k}$ in the first Brillouin zone.

When the dielectric function is independent of $\omega$, the
eigenfrequencies would be easily determined by straightforward
matrix diagonalization \cite{pulos}. However, the dependence of
Fourier elements $\varepsilon_{\vec{G}'}(\omega)$ on the frequency
forces the calculations to be carried out by relying on the
condition of vanishing determinant,
\begin{equation}
det(\mathbf{}M)=0 \; . \label{eq:determinant}
\end{equation}
Numerical calculations are based on finding the zeros of the
determinant of matrix $\mathbf{M}$, as a function of $\vec{k}$ or
$\omega$. The zeros of the complex function is computed using a
least squares method. We have checked the convergence of the
solutions using different initial points.

In the constant dielectric case one chooses a $\vec{k}$ value as
the input and determines the $\omega$ value. This is because
$\omega$ is only on diagonals while $\vec{k}$ is in every element
of the matrix. However, the situation is completely different in
the frequency dependent case. Both $\vec{k}$ and $\omega$ exist in
every element of the matrix. One may solve $\vec{k}$ for the input
values of $\omega$, as well as determining the $\omega$ values
entering the $\vec{k}$ as input. Two methods reveal different
physical pictures \cite{kuzmiak}.

Choosing real $\omega$ values as input, one determines,  in
general, complex $\vec{k}$ values whose imaginary part gives the
spatial attenuation of the propagating wave. On the other hand,
entering real $\vec{k}$ values one solves for complex $\omega$,
whose imaginary part determines the temporal attenuation of the
wave. Since we are mainly interested in the spatial attenuation of
the waves, we followed the first method. However, we checked that
the two approaches give parallel results. As a result of this
procedure we obtain complex wave vector values. We denote the real
and imaginary parts of the wave vector as $k=k_R+ik_I$.

We note that $k_I$ value may imply two different phenomena:
reflection or absorption. If the dielectric function is
real, imaginary part of the wavevector, $k_I$, has a simple
interpretation. The incident wave is totally reflected while
penetrating into the crystal up to a distance of $2\pi/k_I$.
However, if dielectric function is complex one cannot distinguish
between reflection and absorption for a given value of $k_I$. A
mixture of both occurs.

In our computations, we use $11\times11$, parallelogram shaped
grid of plane waves. This corresponds to a $121\times121$
dimensional matrix of elements given in Eq. (\ref{eq:matrix}). We
define the determinant of the matrix as a function of $\vec{k}$
and solve for the zeros of this complex function. We determined
the complex $\vec{k}$ values corresponding to real $\omega$.
Resulting band structures are plotted in Figs.
\ref{fig:directionalgap} and \ref{fig:completegap}.

Although we described our method for the TE modes, TM modes can be
calculated similarly.


\section{Results and Discussion}
\label{sec:results}

When the dielectric function is frequency independent, the Master
equation (\ref{eq:masterTE}) is scalable. That is, the structure of
photonic bands, expressed in terms of scaled frequency
$\omega'=\omega a/2\pi c$, is independent of dimensions of the
unit cell. It only depends on the lattice type and the internal
structure within the unit cell. In the case of a frequency
dependent dielectric, however, such a scaling is not possible, as
a new length scale is introduced into the system. Thus, we first
calculate the band structure with constant $\epsilon$, and then
discuss the change due to frequency dependence of the dielectric
constant.

\begin{figure}
\includegraphics[scale=0.45]{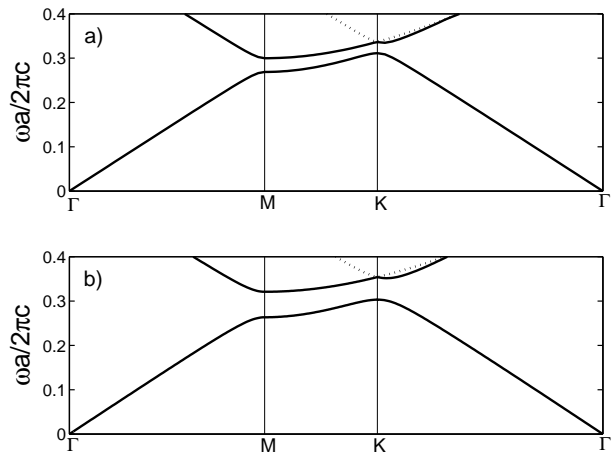}
\caption{TE modes of a triangular vortex lattice with frequency
independent $\epsilon$. Dielectric constants and lattice
parameters are (a) $\epsilon=5.2$ and $a=10\xi$, (b) $\epsilon=8$
and $a=4.5\xi$. Filling fractions of vortices,
$f=(2\pi/\sqrt{3})\times(R^2/a^2)$ with effective radius
$R\simeq2\xi$, are 15\% and 71\%, respectively. Dielectric
constant is the value of dielectric function
(\ref{eq:dielectricfnx}) at the enhancement frequency,
$\epsilon=\epsilon_{\scriptscriptstyle{{\rm loc}}}(\varpi_0)$.
Density profile of the unit cell is treated using the Pad\'{e}
approximation \cite{pade}. (a) There exists a a directional
pseudo-band gap with midgap frequency at
$\omega_g^{\prime}=0.285$. (b) There is a complete band gap with
gap center at $\omega_g^{\prime} = 0.31$.} \label{fig:consteps}
\end{figure}

We give the constant dielectric photonic bands of a triangular
lattice of Rubidium gas for two different sets of parameters, Fig.
\ref{fig:consteps}. The first set, $N=5.5\times10^{20}$
$\text{m}^{-3}$ and $a=10\xi$, has a directional pseudo-gap in the
$\Gamma$M direction, Fig. \ref{fig:consteps}a, when the dielectric
constant is chosen as its value at the enhancement frequency,
$\epsilon=\epsilon_{\scriptscriptstyle{{\rm loc}}}(\varpi_0)=5.2$.
The pseudo-gap lies between the frequencies
$\omega=0.27$-$0.31$($2\pi c/a$), with its center at
$\omega_g=0.285$($2\pi c/a$).

The second set of parameters, $N=6.6\times10^{20}$ $\text{m}^{-3}$
and $a=4.5\xi$, is chosen so that there is a complete photonic
band gap when the dielectric constant is at its enhanced value
$\epsilon=\epsilon_{\scriptscriptstyle{{\rm loc}}}(\varpi_0)=8.0$.
The band gap lies between $\omega=0.30$-$0.32$($2\pi c/a$) with
mid gap frequency $\omega_g=0.31$($2\pi c/a$).

Strong frequency dependence of the dielectric susceptibility, Fig.
\ref{fig:epsilon}, will modify the structure of the bands
significantly. We note that, the dielectric function
(\ref{eq:dielectricfreq}) is different from one only in the
frequency range of $\omega=\Omega_0\pm5\gamma$. The natural
lattice frequency $2\pi c/a$ that we used in the scaling of Fig.
\ref{fig:consteps}, is $7$ orders of magnitude greater than the
decay rate, $\gamma$. For typical values of the lattice parameter
in a rotating BEC, $a\sim 200$ nm, lattice frequency comes is
$2\pi c/a\sim10^{15}$ Hz, whereas the decay rate is only
$\gamma=2\pi\times6\times10^{6}$ Hz. The bands of frequency
dependent $\epsilon_{\scriptscriptstyle{{\rm loc}}}(\omega)$ will
be different than the propagation in vacuum, for only about
$\sim 10 \gamma$ around the enhancement frequency $\Omega_0$. On the
other hand, index enhancement without absorption is achievable in
a more narrow range of frequency, about $0.1\gamma$.

Still, the constant $\epsilon$ bands give us an idea about how to
arrange the lattice parameter, $a$, to obtain a band gap with the
frequency dependent $\epsilon_{\scriptscriptstyle{{\rm
loc}}}(\omega)$. In order to obtain a gap, we must arrange the
enhancement frequency $\Omega_0$ such that, it lies in the band
gap of the corresponding constant dielectric case. A good choice
is to place $\Omega_0$ at the center of the band gap, $\omega_g$.
Thus, we tune the lattice parameter $a$ such that
$\Omega_0=\omega_g^{\prime}(2\pi c/a)$, which gives
\begin{equation}
a=\omega_g^{\prime}\frac{2\pi c}{\omega_{ab}+\varpi_0 \gamma}
\quad , \label{eq:latticepar}
\end{equation}
were $\omega_g^{\prime}$ is obtained from constant dielectric
calculations as in Fig. \ref{fig:consteps}.

In conventional photonic crystals it is generally not possible to
change the lattice parameter, once the sample is manufactured.
However, in the case of a rotating BEC, spacing between the vortex
cores is continuously tunable. Density of the vortices depends on
the rotation frequency, so the lattice parameter $a$ can be
decreased/increased by increasing/decreasing the rotation rate.
Filling factor $f$ of the lattice depends on $f \sim (\xi/a)^2$,
as coherence length $\xi$ determines the vortex core radius (see
Eq. (\ref{eq:pade})). Coherence length $\xi$ can be adjusted by
density $N$. Alternatively, $\xi$ can be adjusted by controlling
$a_{sc}$ via Feshbach resonances \cite{cornish}. We note that, one
might be able to design more convenient sets of parameters for
specific experiments. Using different alkali atoms, like cesium,
stronger index contrasts can be achieved due to larger transition
wavelengths. By employing different index enhancement schemes,
such as the Raman scheme, broader index enhancement windows could
be translated to wider band gaps.

\begin{figure}
\includegraphics[scale=0.45]{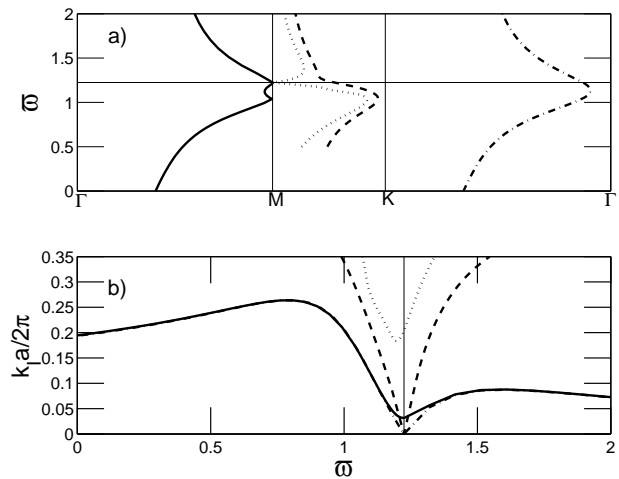}
\caption{ (a) TE modes of triangular vortex lattice with frequency
dependent dielectric function $\epsilon_{\scriptscriptstyle{{\rm
loc}}}(\varpi)$ (Fig. \ref{fig:epsilon}), and (b) imaginary parts
of the wave vector $k_I$ corresponding to each mode. Particle
density is $N=5.5\times10^{20}$ $\text{m}^{-3}$ and lattice
constant is $a=10\xi$. Enhancement frequency $\Omega_0$ is tuned
to the band gap at the M edge ($\omega_g=0.285(2\pi c/a)$) of the
constant dielectric case (Fig. \ref{fig:consteps}a). MK bands are
plotted in a limited region, because of high $k_I$ values out of
the given frequency region. There exists a directional gap in the
$\Gamma$M propagation direction.} \label{fig:directionalgap}
\end{figure}

\begin{figure}
\includegraphics[scale=0.45]{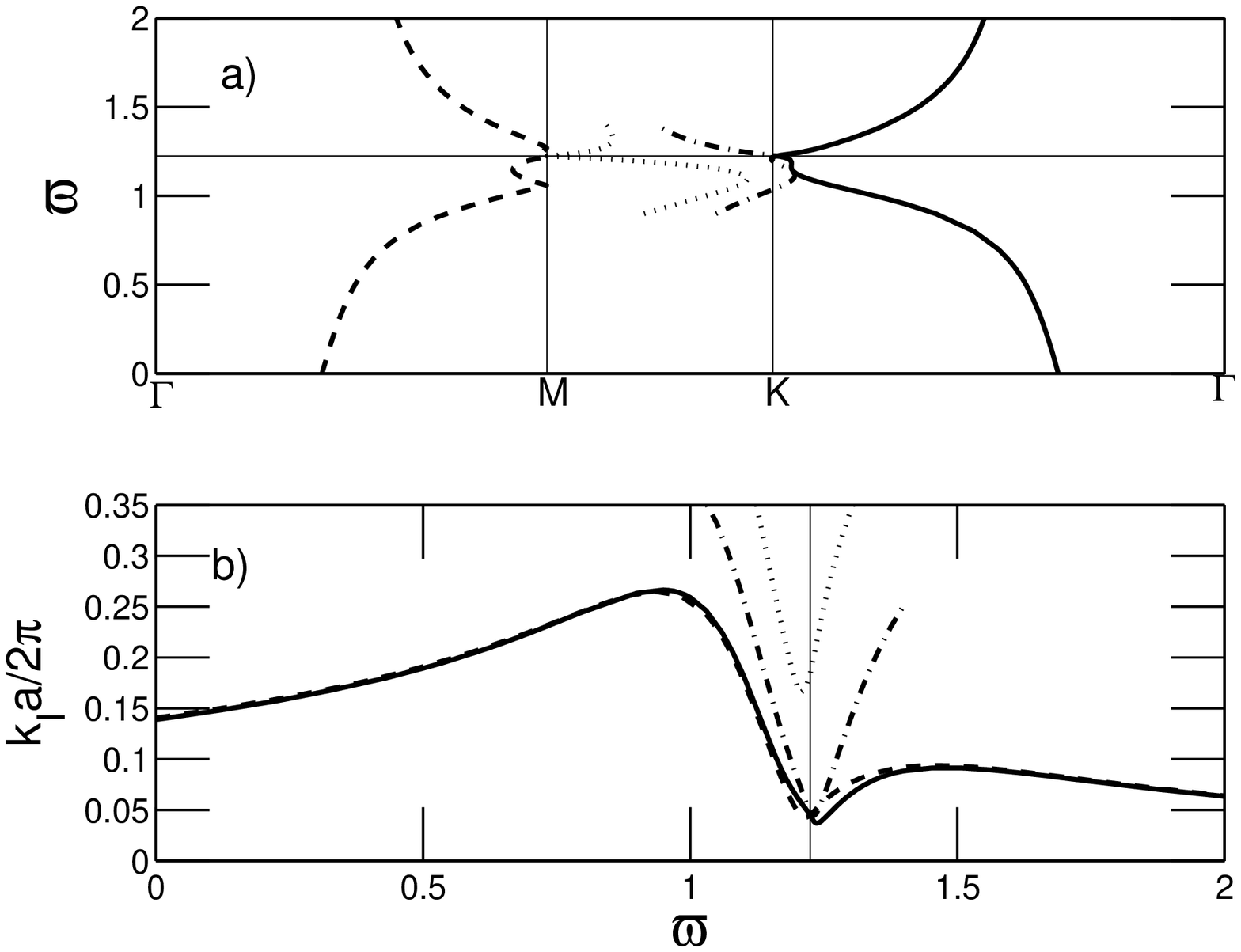}
\caption{(a) TE bands of triangular vortex lattice with frequency
dependent dielectric function $\epsilon_{\scriptscriptstyle{{\rm
loc}}}(\varpi)$ (Fig. \ref{fig:epsilon}b), and (b) imaginary parts
of the wave vector $k_I$ corresponding to each mode. Particle
density is $N=6.6\times10^{20}$ $\text{m}^{-3}$ and lattice
constant is $a=4.5\xi$. Enhancement frequency $\Omega_0$ is tuned
to the band gap at the M edge ($\omega_g=0.31(2\pi c/a)$) of
constant dielectric case (Fig. \ref{fig:consteps}b). There exists
a complete band gap. } \label{fig:completegap}
\end{figure}

Choosing lattice parameter $a$ as in Eq. (\ref{eq:latticepar}), we
calculate the photonic bands for the frequency dependent
dielectric function (\ref{eq:dielectricfnx}). Away from the
enhancement frequency $\varpi_0$, the dielectric function is
complex, (\ref{eq:chi_r},\ref{eq:chi_i}). Band structures are
depicted in figures \ref{fig:directionalgap} and
\ref{fig:completegap} for the same density and filling factor
parameters with figures \ref{fig:consteps}a and
\ref{fig:consteps}b, respectively. In both figures
\ref{fig:directionalgap} and \ref{fig:completegap} the real and
the imaginary parts of the wave vector, $k_R$ and $k_I$, are
displayed separately. Enhancement frequency $\varpi_0=1.22$ is
marked in all plots. The lattice parameters $a$ are chosen as
$a=226$ nm in Fig. \ref{fig:directionalgap} and $a=246$ nm in Fig.
\ref{fig:completegap}.

For a real dielectric function, it is very easy to identify the
band gaps. The wave vector $k$ is real when there is propagation,
and complex (with $k_R$ on the band edge) if the frequency is in a
band gap. For a complex $\epsilon(\omega)$, however,
identification of band gaps is not straightforward. One
can determine the existence of a band gap by considering the
frequency values where $\epsilon$ is real. If a nonzero $k_I$ is
present, then there exists a band gap at that frequency. However,
the width of the band gap cannot be directly identified by
considering only the $k_I$ values away from the enhancement
frequency. For a complex $\epsilon$, nonzero value of $k_I$ may be
due to the absorption as well as the effect of the band gap. Thus,
we first discuss the existence of the band gaps in Fig.
\ref{fig:directionalgap} \& Fig. \ref{fig:completegap} and discuss
the gap widths later.

In Fig. \ref{fig:directionalgap}b, imaginary parts of the wave
vector $k_I$ is plotted for different propagation directions. In
the $\Gamma$M and $\Gamma$K directions only a single band exists
within the enhancement window, while for the MK direction there
are two bands. At the enhancement frequency $\varpi_0=1.22$, two
of these bands have zero $k_I$, while the other two have a complex
wavevector. In accordance with the discussion in the previous
paragraph, we identify the existence of a pseudo-band gap in the
$\Gamma$M propagation direction. Thus, incident light (exactly at
$\varpi_0$) would propagate in the $\Gamma$K, MK directions while
it would be stopped in the $\Gamma$M direction.

In the second case, Fig. \ref{fig:completegap}, all of the four
bands have nonzero $k_I$ at $\varpi=\varpi_0$. This indicates the
existence of a complete band gap at the enhancement frequency.
Incident light is stopped for all propagation directions.

We see that the conclusions for the existence of photonic band
gaps obtained by constant $\epsilon$ calculations are not
modified, even when the strong frequency dependence of
$\epsilon(\omega)$ is taken into account.

Although the existence of directional and complete band gaps are
demonstrated at the enhancement frequency, the widths of these
gaps cannot be determined by only investigating the behavior of
$k_I$. In the vicinity of $\varpi=\varpi_0$, one cannot
distinguish wether the wave is decaying
($e^{-\vec{k_I}\cdot\vec{r}}$) due to absorption, or due to the
existence of a band gap.

To be able to define the width of the gap, we calculate the
behavior of the Poynting vector
\begin{equation}
\vec{S}(\vec{r})=\frac{1}{2}\vec{E}(\vec{r})\times\vec{H}(\vec{r})^*
\quad
\end{equation}
in the crystal. Real part of the Poynting vector,
$\vec{S}_R(\vec{r})$, gives the energy flux of the field at
position $\vec{r}$. Imaginary part $\vec{S}_I(\vec{r})$ is a
measure of the reactive (stored) energy \cite{jackson}. For a
frequency dependent, but real, $\epsilon(\omega)$ Poynting vector
is pure imaginary in the band gaps and real otherwise. However,
for complex $\epsilon(\omega)$ imaginary part of the $\vec{S}$ may
also be due to absorption. Although we make the similar statements
on $\vec{k}_I$ and $\vec{S}_I$, together they are sufficient to
determine the width of the band gap.

Using the results of our band structure calculations, we compute
the average of the Poynting vector, $\langle\vec{S}\rangle$, in
the unit cell. We define
\begin{equation}
\alpha=|\langle S_I\rangle |/|\langle S\rangle | \quad ,
\end{equation}
which corresponds to the rate of reactive energy. Averages of
$\langle S_I \rangle$ and $\langle S\rangle $ are computed along
the $\Gamma$M direction in order to investigate the gap width in
Fig. \ref{fig:directionalgap}.

\begin{figure}
\includegraphics[scale=0.45]{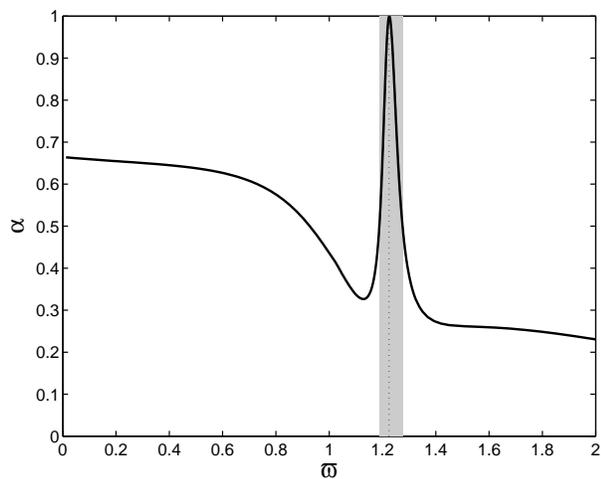}
\caption{Reactive energy ratio $\alpha$ for the $\Gamma M$ band of
Fig. \ref{fig:directionalgap}. Vertical dashed line indicates the
enhancement frequency $\varpi_0=1.22$. Shaded region is the
effective photonic band gap. Width of the peak determines the
width of the gap to be $\omega=\Omega\pm0.043\gamma$ which
corresponds to $\pm1.65$ MHz. } \label{fig:poynting}
\end{figure}

Fig. \ref{fig:poynting} displays a marked increase in the reactive
energy ratio near the enhancement frequency. This increase can not
be directly caused by the imaginary part of
$\epsilon_{\scriptscriptstyle{{\rm loc}}}(\varpi)$, as near the
enhancement frequency this imaginary part is decreasing to zero.
Thus, the peak in the reactive energy ratio must be caused mainly
by the periodicity of the crystal. Presence of the band gap
increases the reactive energy ratio despite  decreasing
absorption.

We define the width of the photonic band gap as the full width at
half maximum of the peak at the enhancement frequency. With this
definition we find an effective gap in the frequency range
$\varpi=\varpi_0\pm0.043$. In familiar units this translates to a
band width of $3.30$ MHz. Using the same method we find a band gap
of width $5.98$ MHz for the parameters of Fig.
\ref{fig:completegap}.

We also performed similar calculations for the TM modes and
obtained similar band structures. TM modes, also, give directional
and complete band gaps when the lattice parameter is properly
tuned. However, since the band gaps of TE and TM modes do not
coincide in general, one can not obtain band gaps for both modes,
without further tuning.

\section{Conclusion} \label{sec:summary}

We calculated the photonic bands for an index enhanced vortex
lattice, considering a frequency dependent dielectric function.
Our motivation was the possibility of the direct measurement of
the rotation frequency in Bose-Einstein condensates using the
directional band gap of the photonic crystal. We validated the
main conclusion of our previous work \cite{mustecap}, that
photonic band gaps can be created via index enhancement on vortex
lattices of BECs. Specifically, we presented two examples showing
that both directional and complete band gaps are possible within
experimentally realizable parameter regimes. For the specific
parameters and index enhancement scheme we considered, band gaps
of order a few MHz width are obtained. We also discussed how band
gaps are designed for specific parameter values, and how band gap
widths can be increased.

Unlike the previous calculations performed for metallic photonic
crystals, complex dielectric function varies rapidly with the
frequency for index enhanced media. Strong frequency dependence is
due to the high dipole moment, established trough atomic coherence
in a narrow frequency  range. To our knowledge, such a periodic
structure composed of index-enhanced media, is investigated for
the first time. We showed that the photonic band structure in such
a medium can be reliably calculated by numerically computing the
zeros of the determinant of the Master equation. We also developed
a method, based on the calculation of Pointing vector, to
determine the effective widths of the photonic band gaps in media
with frequency dependent complex dielectric function.

\acknowledgments \"O.E.M. acknowledges support from a
T\"UBA/GEB\.{I}P grant. M.\"O.O. is supported by a
T\"UBA/GEB\.{I}P grant and T\"UB\.ITAK-KAR\.IYER Grant No.
104T165.

\end{document}